\documentclass[12pt,preprint]{aastex}

\slugcomment{Accepted Ap J Supp.}

\shorttitle{Highest-Energy Photons from EGRET}
\shortauthors{Thompson, Bertsch, O'Neal}

\begin{document}


\title{The Highest-Energy Photons Seen by the Energetic Gamma Ray Experiment Telescope (EGRET) on the Compton Gamma Ray Observatory}


\author{D. J. Thompson\altaffilmark{1}, D. L. Bertsch\altaffilmark{1,2}, and R. H. O'Neal, Jr.\altaffilmark{1,3}}
\altaffiltext{1} {Laboratory for High Energy Astrophysics, NASA Goddard Space Flight Center, Greenbelt, MD, 20771, USA}
\altaffiltext{2} {Retired}
\altaffiltext{3} {Department of Physics, Florida A\&M University, Tallahassee, FL, 32307}
\altaffiltext{3}{NASA-ASSE Faculty Fellowship Program}
\email {djt@egret.gsfc.nasa.gov,dlb@egret.gsfc.nasa.gov, oneal@milkyway.gsfc.nasa.gov}

\clearpage

\begin{abstract}
During its nine-year lifetime, the Energetic Gamma Ray Experiment Telescope (EGRET) on the Compton Gamma Ray Observatory (CGRO) detected 1506 cosmic photons with measured energy E$>$10 GeV.  Of this number, 187 are found within a $1^\circ$ of sources that are listed in the Third EGRET Catalog \citep{hart99} and were included in determining the detection likelihood, flux, and spectra of those sources.  In particular, five detected EGRET pulsars are found to have events above 10 GeV, and together they account for 37 events.  A pulsar not included in the Third EGRET Catalog has 2 events, both with the same phase and in one peak of the lower-energy gamma-ray light-curve.  Most of the remaining 1319 events appear to be diffuse Galactic and extragalactic radiation based on the similarity of the their spatial and energy distributions with the diffuse model (\citet{hunt97,kum98}) and in the E$>$100, MeV emission.  No significant time clustering which would suggest a burst was detected.  
\end{abstract}




\section{Introduction}

Between 10 GeV and about 100 GeV lies a largely unexplored portion of the electromagnetic spectrum.  Neither satellite gamma-ray telescopes nor ground-based Cherenkov gamma-ray telescopes have had significant sensitivity in this energy band. Nevertheless, it is an astrophysically interesting region.  Some sources are seen that appear on both sides of the gap (some BL Lac objects and at least one supernova remnant); other sources seem to vanish in this region (Flat Spectrum Radio Quasars, pulsars, unidentified sources).  Particularly in preparation for the upcoming GLAST (Gamma Ray Large Area Space Telescope) mission that will be studying this band, any insight into this portion of the spectrum should prove valuable. 

One source of data above 10 GeV is the EGRET database.  EGRET was planned for a relatively short mission in which few photons above 10 GeV were anticipated; therefore it was not optimized for these high energies.  In particular, EGRET used a monolithic anticoincidence detector to screen out charged particles (See \citet{hughes80,kanb88,kanb89,thom93}, and \citet{espos98} for a description of the instrument and its capabilities.), but very-high-energy gamma rays interacting in the instrument often produced secondary particles that scattered into this anticoincidence detector, causing a self-veto. Due to this effect, the high energy sensitivity was reduced so that the  effective area at 10 GeV was about 50\% of its peak sensitivity between 500 and 1000 MeV \citep{thom93}. With the extended lifetime of the CGRO mission, however, even the limited EGRET sensitivity produced over 1500 photons with measured energies greater than 10 GeV.

This paper is a systematic study of these highest-energy photons from EGRET by members of the original EGRET team.  We reviewed the original analysis of all the candidate high-energy photons to assure that the event structuring from the spark chamber images was correct and  consistent selections were applied.  Although much of the subsequent analysis was done on the individual photons, we also extrapolated the EGRET performance parameters to higher energies than were covered by the EGRET calibration, in order to determine exposure values. 

This work is not the first to use some of these data.  Some of the events are located within the regions of sources that are listed in the EGRET catalogs, the most recent of which is the Third EGRET Catalog \citep{hart99}.  These were included in assessing the source significances, their locations and their energy spectra.  The diffuse Galactic spectrum \citep{hunt97} and the diffuse extragalactic radiation \citep{kum98} included data points above $10$ GeV.  One of the BATSE gamma ray bursts seen on February 17, 1994 was also detected by EGRET, and among the photons it observed was one event with energy $E = 18.4$ GeV.  Its arrival was delayed by 79 minutes from the time of the burst \citep{hur94}.

Some archived EGRET data above 10 GeV have been analyzed previously (e.g. \citet{ding01,fons01,strong04}).  Although the archive does not have the final screening that was applied to the current data set, the differences are small.  Some of the results presented here duplicate these earlier analyses.

\section{EGRET Data Analysis}

EGRET was the high-energy gamma-ray telescope on the Compton Gamma Ray Observatory (CGRO), whose mission lasted from 1991$-$2000.  The telescope covered the energy range from about 30 MeV to over 20 GeV, limited at the high end by statistics.   EGRET recorded gamma-ray photons individually as electron-positron pair production events, which were processed automatically (with manual verification) to provide the arrival direction and energy of each photon.  The point-spread function was energy-dependent, having a 67\% containment radius of approximately $5.8^\circ$ at 100 MeV and smaller values at higher energies. This radius was $0.5^\circ$ at 10 GeV, the highest energy for which experimental calibration was done \citep{thom93}.  The arrival time of each gamma ray was recorded in Universal Coordinated Time (UTC) to an accuracy of better than $100 {\mu}s$. The field of view of EGRET extended to more than $30^\circ$ from the instrument axis, although the sensitivity at angles beyond $30^\circ$ was less than 15\% of the on-axis sensitivity.  Because of the low flux level of the high energy gamma rays, observing periods were typically 2$-$3 weeks.

The EGRET database includes over $5\times 10^6$ gamma-ray events that were recorded during the 9 year mission (Note that EGRET was not operated continuously due to Earth occultation, South Atlantic Anomaly passages, and spark chamber gas conservation in the second half of the mission.). Some of these events have arrival direction that exceed $40^\circ$ from the instrument axis, regarded as the maximum field of view, or whose arrival may intercept the Earth's limb, known to be a bright gamma-ray source \citep{thom81}.  Moreover some of the events were recorded during periods of testing or calibration or non-standard instrument configuration.  For the study of the very high energy events, a selection was made for energies $E>10$ GeV and for valid observing times.  Then additional cuts were made to select events whose arrival direction are within $40^\circ$ of the instrument axis and whose arrival direction relative to the Earth's zenith is less than $105^\circ$.  This angle was selected to be 4 standard deviations away from the geometric atmospheric boundary, given the instrument point-spread-function at these high energies.  A total of 1550 events meeting these criteria were found.

Spark chamber images of each of the selected events were reviewed by an experienced analyst to eliminate events whose structuring was questionable, or whose initial interaction point might possibly have been within the spark chamber walls. At these high energies, tracks in the EGRET spark chamber are expected to be straight, with little scattering.  Events without at least one straight track were eliminated.  The energy measurement depends on the electron-positron pair depositing energy in the energy-measuring calorimeter, the Total Absorption Shower Counter (TASC); events in which the tracks did not point toward the calorimeter were eliminated.  On the basis of such screening, only 44 events were eliminated from the sample, leaving 1506 high energy events.  Consequently, this additional review has little effect on earlier analyses, in particular the galactic \citep{hunt97} and extra-galactic
\citep{kum98} papers.  The event associated with the February 17, 1994 burst remains in the list of valid events.  A table listing the arrival date and time, energy and energy uncertainty, arrival direction in galactic coordinates, energy class, and angle from instrument axis of each photon with energy $E>10$ GeV is available in FITS format in the anonymous FTP site on {\it gamma.gsfc.nasa.gov} in the directory {\it /pub/VHE\_events} named {\it VHE\_event.tbl}.   In this table, energy class A signifies a good quality of measurement by the TASC with both of the pair tracks pointed toward the interior portion of the TASC, and energy class B signifies a poorer quality in which some energy leakage out of the side of the TASC probably occurred. For conciseness, we refer to the $E>10$ GeV EGRET events as VHE events elsewhere in this paper. In astrophysics, VHE typically refers to data from ground-based atmospheric Cerenkov telescopes at energies above 100 GeV.  Our use of the same term emphasizes that these highest-energy EGRET events begin to bridge the gap between space-based and ground-based telescopes. 

\section{Studies of the High-Energy Data}

\subsection{Basic Information about the Data Set}

Table 1 shows the distribution of measured event energies for the E $>$ 10 GeV data. Energy measurements from the EGRET TASC calorimeter (8 Radiation Lengths of NaI) suffer from substantial shower leakage above 10 GeV, so that the measurement uncertainty grows with increasing energy. The FWHM on energy measurement is about 25\% at  10 GeV, the highest energy for which experimental calibration was done \citep{thom93}.  At the highest energies, the uncertainty in individual photon energy measurements is estimated to be typically 50\%. Examples of energy uncertainties for individual gamma rays are shown in Tables 5 - 9. As expected in almost any astrophysical situation, the number of events decreases rapidly with increasing energy.  In the case of EGRET, this fall-off is exacerbated by the fact that the EGRET detection efficiency also falls off with energy above 10 GeV.

Table 2 shows the distribution of arrival directions of VHE events within EGRET, expressed as an angle from the vertical axis (pointing direction of EGRET). Because the tracks in the spark chamber show little scattering, this distribution essentially reflects the geometric response of the instrument.  As the angle from the axis increases, the solid angle grows, but the projected area of the triggering telescope decreases.  Over 60\% of the photons were detected at arrival angles between $10^\circ$ and $25^\circ$ and only 4\% were detected above $30^\circ$.

\subsection{All-Sky Distribution of the $E>10$ GeV Events}

EGRET exposure maps generally were made for events within $30^\circ$ of the instrument axis.  An all-sky exposure map for energies above 4 to 10 GeV using that angle cutoff is part of the standard EGRET products.  For the purposes of making an all-sky map that is weighted by exposure, 59 events beyond $30^\circ$ were removed from the list in order to be consistent with all-sky exposure map.  The EGRET effective area decreases above 1 GeV due to self-veto by secondaries generated when high-energy gamma rays shower in the calorimeter. The instrument response was not directly calibrated above 10 GeV.  However, Monte Carlo modeling of the self-veto \citep{thom93, kum98} showed that the effective area above 10 GeV decreases approximately exponentially with energy as $exp(-E/36 \,GeV)$.  This function, when weighted by a power-law spectrum gives effective areas relative to the effective area for 4 to 10 GeV shown in Table 3 for three specific power-law indices.  Since the galactic diffuse emission is reasonably fit with a power-law of spectral index 2.1 (\citet{hunt97}), the 4 to 10 GeV all-sky exposure map was scaled by 0.645, and then the ratio of the counts map to the resulting exposure map was generated to produce the map shown in Fig. 1. Although this extrapolation of the response inherently introduces additional systematic uncertainty \citep{hunt97, kum98}, the photon statistics for any individual sources are so small above 10 GeV that the overall uncertainty in any calculated VHE flux is dominated by the statistics.  

While strictly speaking, this map is an intensity map, the events are so sparse that it is better to regard it as a relative counts map weighted by the exposure.  The character of this map is similar to EGRET all-sky maps at lower energies with the galactic plane being one of its most striking features, and that indicates that much of the emission is due to galactic diffuse processes and galactic sources.  In Fig.2, the galactic latitude and longitude distributions of the VHE photons are compared with the corresponding distributions for the $>$100 MeV energy range.  The similarity in both latitude and longitude between the two ranges shows that the VHE photons have essentially the same source characteristics as the $>$ 100 MeV photons where diffuse processes are dominant(\citet{hunt97}).

\subsection {$E>10$ GeV Events in Third EGRET Catalog Sources}

A search was made to identify the VHE events that fall within $1^\circ$ of EGRET 3rd Catalog sources.  This search radius was chosen to account for the average source uncertainty as well as the uncertainty in the location of the VHE events ($0.5^\circ$).  A total of 187 events were identified, and these are listed in Table 4.  The last column in Table 4 gives the number of photons above 10 GeV expected within the  $1^\circ$ circle from the diffuse Galactic and extragalactic emission.  To derive these values, the diffuse models (\citet{hunt97,kum98}) for the 4 to 10 GeV energy interval were multiplied by the summed exposure map for the mission in that same energy band.  The resulting diffuse counts map was then renormalized to have the same total count as the observed number of $>10$ GeV photons (1506).  For any source location, a Poisson probability of chance occurrence can be calculated using the number derived from this map as the average. The use of a renormalized counts map rather than an absolute calculation is critical, because the EGRET diffuse model is known to under-predict compared to the observations above 1 GeV (the ``GeV excess''). The renormalization essentially eliminates the spectral discrepancy. More recent diffuse models, such as \citet{strong04}, provide better absolute fits to the observed spectrum, but the large-scale spatial distribution does not differ dramatically from the earlier work.  At energies above 10 GeV the present statistics are far too small to distinguish such models based on spatial distributions. Use of a more recent model, similarly normalized to the observed number of photons, would not, therefore, have any significant impact on the diffuse estimates in Table 4.

Some obvious sources in Table 4 are the bright EGRET pulsars: Crab (3EG J0534+2200), Geminga (3EG J0633+1751), Vela (3EG J0834$-$4511), and PSR B1706$-$44 (3EG J1710$-$4439). Comparing the number of photons seen with those expected from the diffuse emission shows that essentially all events are likely to be associated with the sources. As a confirmation, a likelihood analysis was carried out for the pulsar locations, using the standard EGRET software. (The EGRET likelihood analysis is described in detail elsewhere \citep {mat96}.) The analysis reproduces the pulsars as significant detections:  Crab (8.5 $\sigma$), Geminga (7.1 $\sigma$), Vela (4.8 $\sigma$), and PSR B1706$-$44 (6.1 $\sigma$). These pulsars are steady, Galactic sources which received substantial EGRET exposure and have relatively flat energy spectra. Their appearance above 10 GeV is perhaps not surprising, except for the fact that Vela and Geminga in particular have spectra with distinct spectral breaks in the few GeV range.  The fact that the spectra extend above 10 GeV suggests that these breaks are not sharp cutoffs.  The statistics are, nevertheless, too small to define a detailed spectral shape at these high energies.  

Other clusters of events near 3EG sources are near the Galactic Plane in the high-intensity region toward the inner part of the Galaxy.  In this region, the diffuse Galactic emission is sufficiently high that the probability of chance coincidence is not negligible.  A cluster of 8 photons is seen, for example, within $1^\circ$ of 3EG J1746$-$2851 near the Galactic Center, but the diffuse emission expected in such a circle can account for 3.5 of those photons. The excess is significant (3\% chance probability; there is no trials factor in the probability calculation, because the source location is already defined), but the diffuse emission cannot be completely neglected. In addition to the source just mentioned, the highest-confidence unidentified 3EG sources for having E$>$10 GeV emission are 3EG J1410$-$6147, 3EG J1627$-$2419, 3EG J1655$-$4554, 3EG J1714$-$3857, 3EG J1837$-$0606, and 3EG J1856+0114, all with less than 10\% probability that the number of photons seen is a fluctuation on the diffuse emission. 

One, and only one, of the blazars seen by EGRET shows more than three nearby VHE photons. PKS 1406$-$076 has four photons within $1^\circ$.  The probability that these four come from the diffuse emission is less than 10$^{-6}$. These photons appear in two pairs.  The first pair occurred close to the time of the bright flare of this source seen in January, 1993, CGRO Viewing Periods 206 and 207.  The second pair occurred in December, 1994, during Viewing Period 405, at a time when the E$>$100 MeV emission from this blazar was not strong.  It appears as only an upper limit in the EGRET catalog.  
Clusters of three photons (all statistically improbable based on the expected diffuse emission) are found near three other blazars:  0430+2859 (0.04 chance probability), 1622$-$297 ($<$ 0.01 chance probability), and 2155$-$304 ($<$ 10$^{-5}$ chance probability).  A more detailed review of the VHE photons that might be associated with blazars is in progress \citep{ding04}. 

\subsection{Search for Spatial Clustering of the $E>10$ GeV Events}

In order to search for high energy source locations other than the ones associated with cataloged EGRET sources, the high energy photon events above 10 GeV were surveyed for spatial clustering. Test circles with $1^\circ$ radius were stepped in $0.5^\circ$ increments over the entire sky and the number of VHE events was recorded for each position.  Each location of three or more photons occurring within the error radius ($1^\circ$) were processed using the EGRET likelihood analysis program in order to determine whether such events represent gamma ray emission significantly greater than that expected from diffuse background.  All but one of the clusters turned out to be consistent with known 3EG sources. 

One unidentified five photon cluster was found between galactic longitudes 179 and 180, and galactic latitudes $-$1 and $-$2.  Using   likelihood analysis to localize this excess produces the map shown in Fig. 3.  The cluster direction lies near the center of the shell of supernova remnant S147 (G180.0$-$1.7), which is $3^\circ$ in diameter (see summary by \citet{gre04}). The cluster is within $1^\circ$ of PSR J0538+2817, which is associated with this supernova remnant (see summary information at {\it www.atnf.csiro.au/research/pulsar/psrcat)}.  Treated as a single trial, the statistical significance of this cluster is 4.8$\sigma$, but it was found in the course of a search over all VHE photons.  Likelihood analysis of this direction at energies lower than 10 GeV in the EGRET maps produces no significant signal. The likelihood analysis shows that PSR J0538+2817 lies outside the 99\% confidence contours, making it unlikely that this pulsar is the source of these photons. These photons were detected within the seven years of the mission between 1992 and 1999, with no two photons occurring less than approximately a year apart.  Although we cannot rule out the possibility of a real source with an extremely flat spectrum, the data appear consistent with a fluctuation in the diffuse emission. Archival data from RXTE, ASCA and Chandra show no X-Ray emission from this location other than that associated with PSR J0538+2817.

\subsection{Pulsar Timing Analysis}

Timing analysis using the standard EGRET pulsar programs was carried out for the VHE photons close to known EGRET pulsars: Crab, Geminga, Vela, PSR B1706$-$44, and PSR B1951+32.  No VHE photons were found close to PSR B1055$-$52.  For Geminga, the timing parameters were those of Mattox, Halpern, and Caraveo (1998).  For the others, data were obtained from the CGRO Timing Data Base ({\it http://pulsar.princeton.edu/ftp/gro/psrtime.dat}) and the online radio timing information at the Australian Pulsar Timing Archive\\ 
({\it http://www.atnf.csiro.au/research/pulsar/archive/}) and Jodrell Bank Timing Database\\ ({\it http://www.jb.man.ac.uk/research/pulsar/}).  

The results for VHE photons within $1^\circ$  of five pulsars are shown in Fig. 4 and Tables 5 through 9, ordered by increasing Right Ascension. Even though the pulsar positions are known, the use of a $1^\circ$ radius is useful, because the sources are sufficiently bright that the background is not significant.  The increased statistics from adding in more of the point spread function increases the significance of the pulsed detection. The VHE photon phases, shown in 0.05 phase bins, are compared to the light curves (dotted lines) seen at lower gamma-ray energies (see summary by Thompson (2004)). Although the statistics are small, there is evidence of E$>$10 GeV pulsed emission for most of these pulsars, with the possible exception of the Crab.  Some notes on the individual pulsars:

Crab. The pulsed emission covers approximately 0.6 of the full phase. Of the 10 VHE photons, 7 appear in the pulsed phase region and 3 in the unpulsed region. The statistics do not, therefore, warrant a claim of pulsed emission, nor do they rule it out.  As discussed, for example, by \citet{dej96}, an unpulsed inverse Compton component from the Crab Nebula is expected to become dominant at these energies.  The fact that 7 of the 10 photons have measured energy above 20 GeV is an indication of a very flat spectrum for this emission compared to that of any of the other pulsars.  Only five photons in the entire EGRET database have measured energies above 100 GeV, although with very large uncertainty. Two of these photons are associated with the Crab, including the single highest-energy cosmic photon observed by EGRET. Both have phases close to that of the first pulse in the Crab light curve. The Crab will continue to be a target of great interest for both GLAST and ground-based gamma-ray telescopes. 

Geminga.  The VHE events are clearly concentrated in the pulsed portion of the phase plot, with a peak associated with the second pulse seen at lower energies. Five of the 10 photons are consistent with the second pulse. This trend toward a stronger second peak was noted previously with EGRET data above 5 GeV \citep{thom01}.

Vela.  Only four VHE events are seen, but all are in the pulsed portion of the light curve, and two of the four are within the second gamma-ray pulse. Three of the four events were seen in a space of three days, although this observation was not the deepest exposure to Vela.  

PSR B1706$-$44.  A clear concentration of VHE events in the pulsed portion of the phase plot is seen, with 5 of the 9 photons associated with the first pulse and 2 additional photons consistent with the second pulse.  This result agrees with the E$>$5 GeV observations (which include the same photons).

PSR B1951+32.  Although only two VHE photons are seen from the direction of this source, both have phase associated with the  second pulse from the lower-energy data.  These two have phases separated by less than 0.01.  Despite the minimal statistics, we argue that these two photons do constitute evidence for pulsed emission above 10 GeV.  

An earlier analysis based on archived EGRET data \citep{fons01} reached similar conclusions. 

\subsection{Search for $E>10$ GeV Photons Associated with Gamma Ray Bursts}


The BATSE burst seen on February 17, 1994 in which EGRET detected delayed gamma ray emission with one event of energy 18.4 GeV arriving 79 minutes from the time of the burst \citep{hur94} prompted a search for delayed VHE emission from other BATSE bursts.  For each burst in the BATSE catalog\\ ({\it www.batse.msfc.nasa.gov/batse/grb/catalog/current/tables/basic\_table.txt}) the list of VHE events was searched to identify any events that were within $5^\circ$ of the BATSE direction and within 2 hours following the detection time.  The choice of $5^\circ$ was governed by the BATSE location uncertainty.  The only event that was found was for the February 17, 1994 burst.

A time clustering search was also made independent of the BATSE catalog.  Time differences (t) between successive events were formed and their probability of occurrence was calculated given the mean VHE rate per live-day (m) according to $1-exp(-t*m/86400)$.  Pairs of events with probabilities less than 0.01 were examined to see if their arrival was consistent with the same sky location.  No pairs were found.

\section{Conclusions}

Despite the limited statistics, the EGRET data offer hints of the astrophysical information available at photon energies between 10 and 100 GeV. The agreement in overall celestial distribution at these energies with that seen at lower energies demonstrates that a pair production telescope can identify such high energy photons even at the limits of its resolution. The detection of VHE emission from pulsars provides clear evidence that these spectra extend into this energy range. Indications of E$>$10 GeV radiation from unidentified Galactic sources and a few blazars illustrate the prospects for important discoveries with the GLAST Large Area Telescope, as well as with the new generation of ground-based gamma-ray observatories whose energy threshold is being pushed downward.  

\acknowledgments

The event display software used for this analysis was provided by Toby Burnett, whose help we are appreciate. We are grateful to the many scientists, engineers, technicians, managers, and analysts who contributed to EGRET.  We thank the radio astronomers who constructed and maintained the CGRO pulsar database.



\clearpage

\begin{figure}          
\plotone{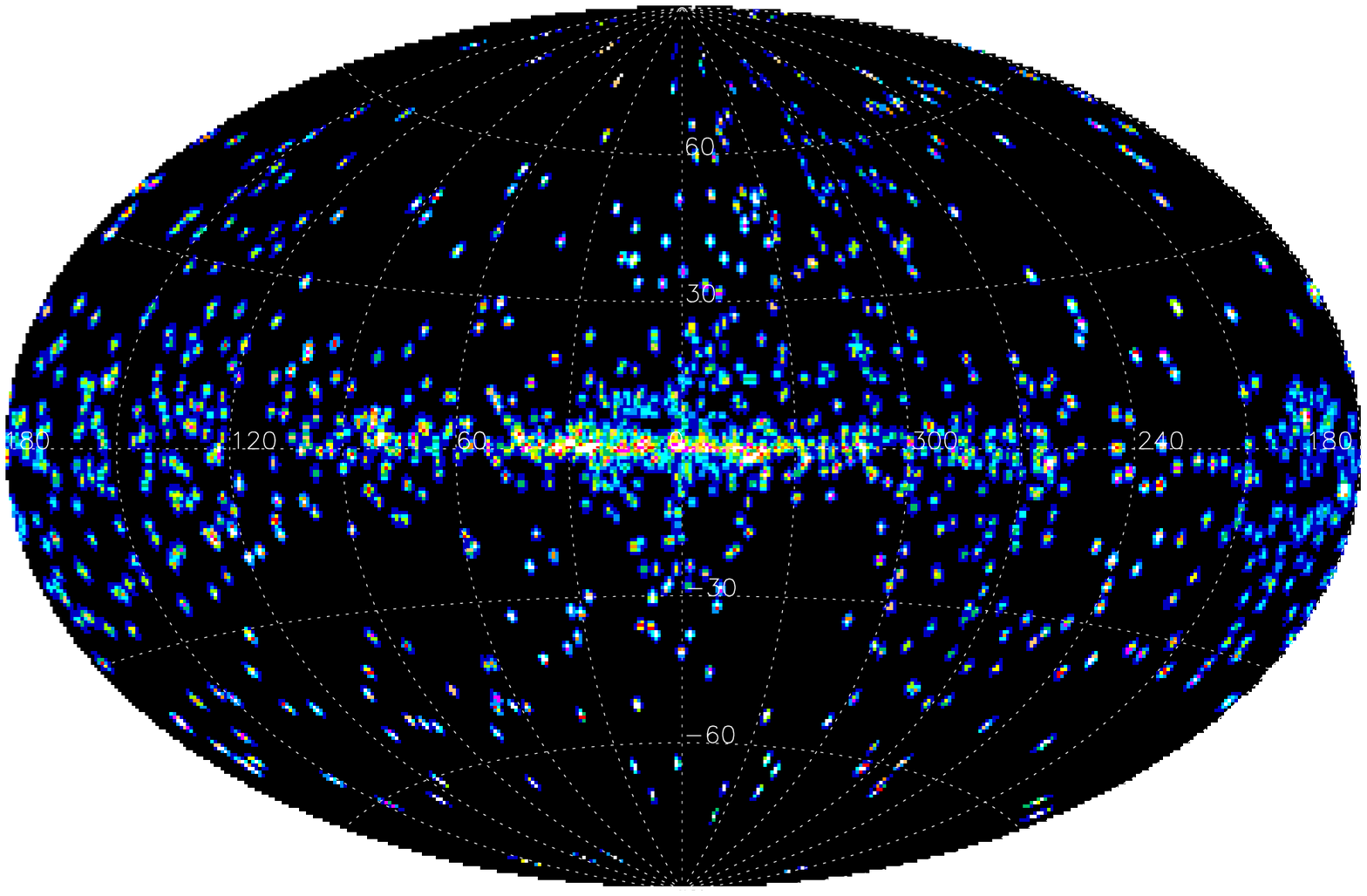}
\caption{Intensity map of EGRET events with energy above 10 GeV.
\label{fig1}}
\end{figure}

\clearpage

\begin{figure}         
\epsscale{0.75}
\plotone{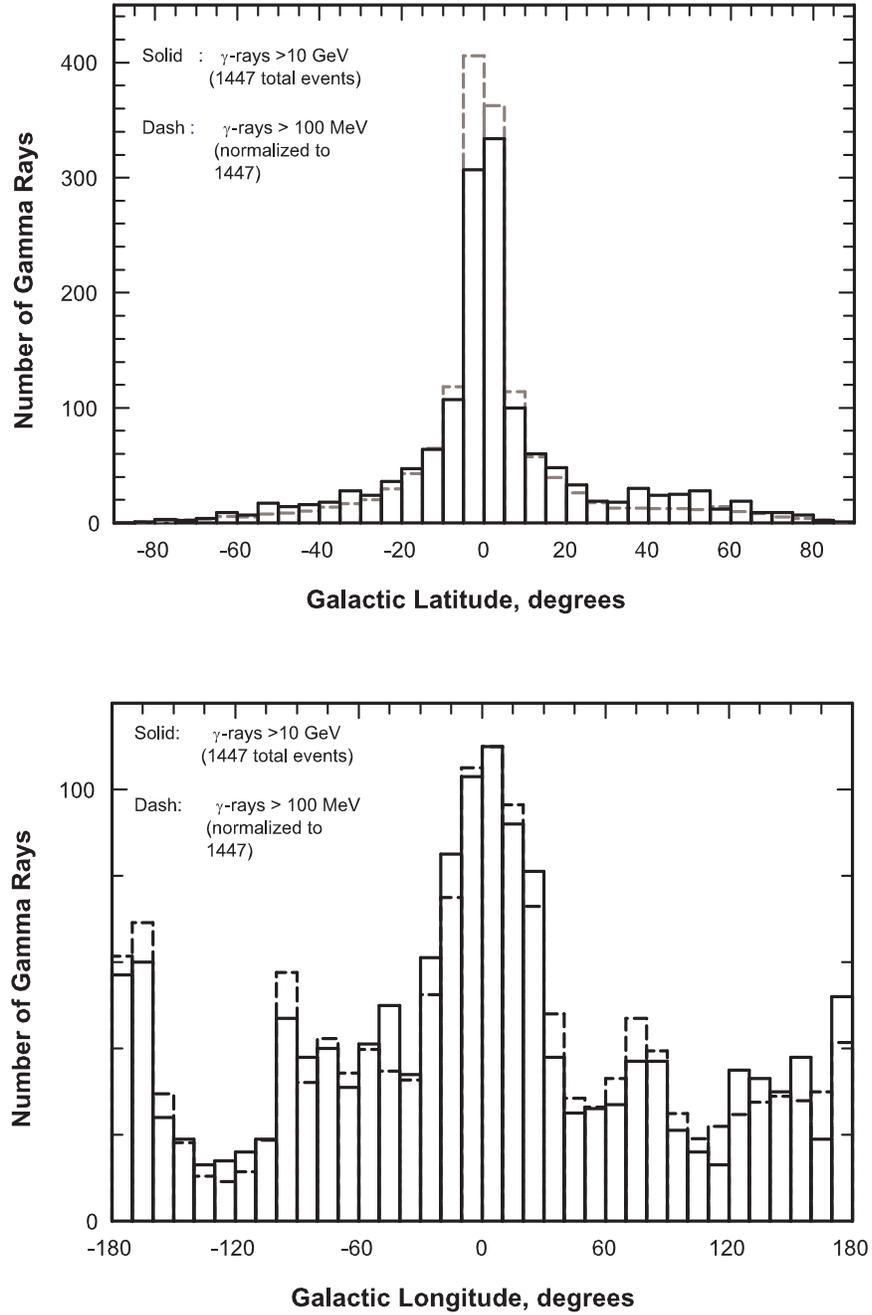}
\caption{$E>10$ GeV Galactic latitude and longitude distributions compared with the same distributions for energy $>$100 MeV.\label{fig2}}
\end{figure}

\clearpage

\begin{figure}         
\plotone{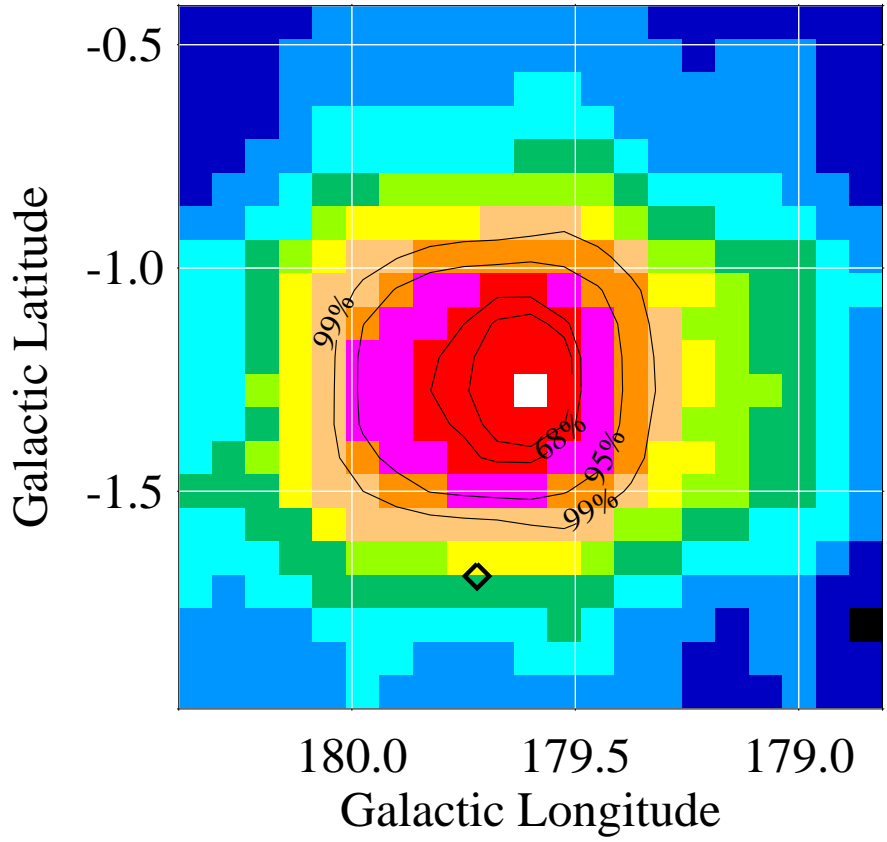}
\caption{Maximum likelihood localization of an unidentified cluster of VHE events, showing probability contours of 50\%, 68\%, 95\%, and 99\%.  The diamond that marks the location of PSR J0538+2817 is outside the 99\% contour.\label{fig3}}
\end{figure}

\clearpage
\begin{figure}         
\hspace{0.4in}
\plotone{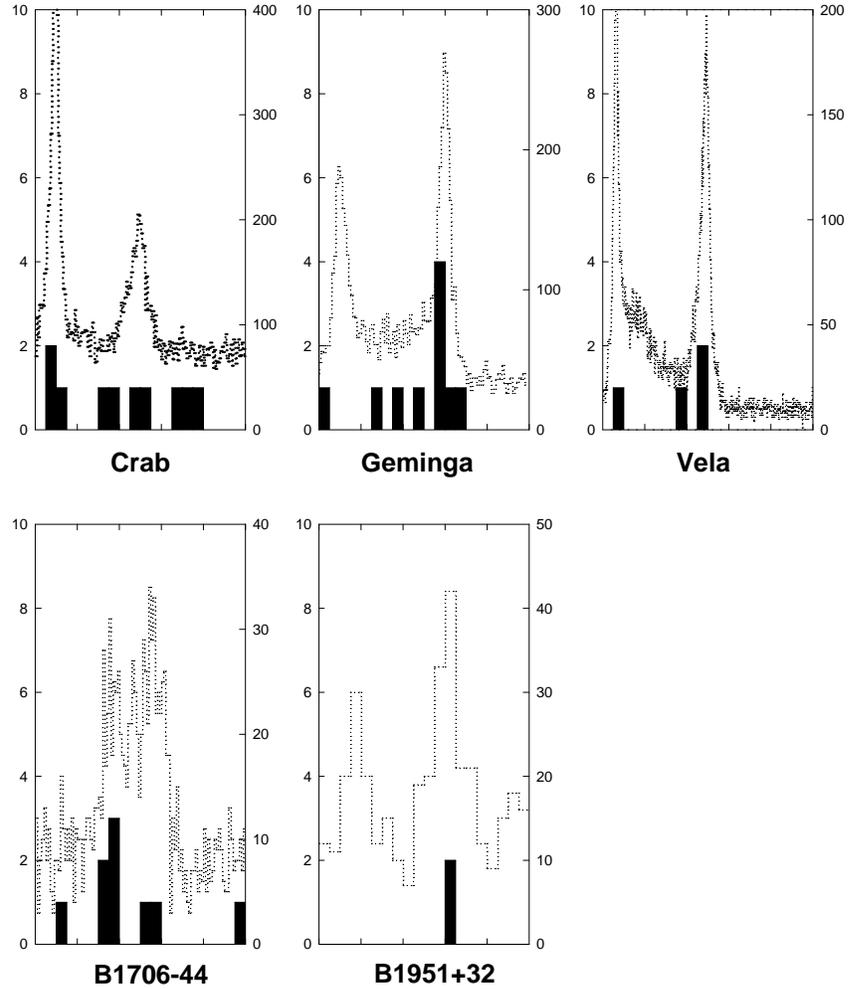}
\caption{Light curves of five gamma-ray pulsars, comparing VHE events with lower-energy data. Dotted curves: light curves from EGRET data above 100 MeV (Thompson, 2004), right scale. Solid histogram: photons above 10 GeV within $1^\circ$ of the pulsar, left scale. Each plot shows the full phase range for a full rotation of the neutron star. 
\label{fig4}}
\end{figure}

\clearpage





\clearpage

\begin{deluxetable}{lr}
\tabletypesize{\footnotesize}
\tablecaption{Energy Distribution of $E>10$ GeV Events. \label{tbl-1}}
\tablewidth{0pt}
\tablehead{
\colhead{Energy Range} & \colhead{Number}\\
  \colhead{(GeV)} &  }
\startdata
10 $-$ 15 &   787 \\
15 $-$ 25 &	455 \\
25 $-$ 50 &	211\\
50 $-$ 100&	48\\
$>$ 100 &	5
 \enddata

\end{deluxetable}

\clearpage

\begin{deluxetable}{rr}
\tabletypesize{\footnotesize}
\tablecaption{Angular Distribution of $E>10$ GeV Events in EGRET. \label{tbl-2}}
\tablewidth{0pt}
\tablehead{
\colhead{Angle Range} & \colhead{Number}\\
  \colhead{(Deg)} &  }
\startdata
0 $-$ 5 &   150 \\
5 $-$ 10 &	301 \\
10 $-$ 15 &	323\\
15 $-$ 20 &	302\\
20 $-$ 25 &	239\\
25 $-$ 30 &	132\\
30 $-$ 35 &	36\\
35 $-$ 40 &	23
 \enddata

\end{deluxetable}

\clearpage


\begin{deluxetable}{rrp{0.5in}rrr}
\tabletypesize{\footnotesize}
\tablecaption{Effective Area Relative to the 4 to 10 GeV Value\label{tbl-3}}
\tablewidth{0pt}
\tablehead{
        \multicolumn{2}{l}{\bf Energy} & &
        \multicolumn{3}{c}{\bf Relative Eff. Area} \\[-0.02in]
           \multicolumn{2}{l}{\bf Range (GeV)}\\[-0.02in]           &  &  & \multicolumn{3}{c}{\bf Spectral Index} \\[-0.02in]
           \colhead{\bf Low} & \colhead{\bf High} &  &  \colhead{\bf 1.80} &                       \colhead{\bf 2.10} & \colhead{\bf 2.40}}
         
\startdata
        10 &    20                      & & 0.807 & 0.808 & 0.809 \\[-0.02in]
        10 &    50                      & & 0.683 & 0.706 & 0.723 \\[-0.02in]
        10 &    70                      & & 0.649 & 0.678 & 0.702 \\[-0.02in]
        10 &   100                      & & 0.616 & 0.655 & 0.678 \\[-0.02in]
        10 &   120                      & & 0.602 & {\bf 0.645} & 0.681 \\[+0.02in]
        20 &    50                      & & 0.514 & 0.512 & 0.528 \\[-0.02in]
        20 &    70                      & & 0.464 & 0.478 & 0.491 \\[-0.02in]
        20 &   100                      & & 0.421 & 0.443 & 0.463 \\[-0.02in]
        20 &   120                      & & 0.403 & 0.429 & 0.452 \\[+0.02in]
        50 &    70                      & & 0.233 & 0.234 & 0.234 \\[-0.02in]
        50 &   100                      & & 0.183 & 0.186 & 0.189 \\[-0.02in]
        50 &   120                      & & 0.164 & 0.169 & 0.173 \\[+0.02in]
        70 &   100                      & & 0.120 & 0.121 & 0.121 \\[-0.02in]
        70 &   120                      & & 0.102 & 0.104 & 0.105 \\[-0.02in]
\enddata
\end{deluxetable}

\clearpage

\begin{deluxetable}{rrrrrrrrrr}
\tabletypesize{\footnotesize}
\tablecaption{$E>10$ GeV Events Near 3EG Sources. \label{tbl-4}}
\tablewidth{0pt}
\tablehead{
\colhead{Source Name} &\colhead{Source}&\colhead{Source}&\colhead{Event}&\colhead{Event}&\colhead{Energy}&\colhead{Angle}&\colhead{Event}&\colhead{Event}&\colhead{Expected}\\
  &\colhead{Long.} &\colhead{Lat.}&\colhead{Long.}&\colhead{Lat}&\colhead{(MeV)}&\colhead{(Deg)}&\colhead{Date}&\colhead{UT}& \colhead{Diffuse}}
\startdata
   J0210$-$5055     & 276.10& -61.78&  274.07& -61.59&  14070.&  0.981&  09/25/91&  13:38:55&   0.036\\
   J0210$-$5055     & 276.10& -61.78&  275.70& -61.83&  10749.&  0.198&  03/17/96&  08:59:24&   0.036\\
   J0222+4253     & 140.14& -16.77&  139.73& -16.57&  12243.&  0.440&  10/21/98&  16:50:40&   0.056\\
   J0348+3510     & 159.06& -15.01&  158.55& -15.51&  11576.&  0.699&  05/03/00&  14:21:02&   0.135\\
   J0429+0337     & 191.44& -29.08&  192.10& -29.73&  16071.&  0.870&  04/24/00&  13:50:01&   0.097\\
   J0433+2908     & 170.48& -12.58&  170.27& -12.16&  29178.&  0.463&  02/28/97&  03:52:28&   0.320\\
   J0433+2908     & 170.48& -12.58&  170.40& -12.59&  14146.&  0.082&  08/15/96&  01:25:44&   0.320\\
   J0433+2908     & 170.48& -12.58&  170.45& -12.68&  15749.&  0.101&  08/20/95&  07:31:17&   0.320\\
   J0439+1105     & 186.14& -22.87&  185.52& -22.85&  29308.&  0.569&  03/25/93&  14:27:34&   0.157\\
   J0442$-$0033     & 197.20& -28.46&  197.65& -28.56&  11367.&  0.408&  08/25/94&  16:37:21&   0.085\\
   J0500+2529     & 177.18& -10.28&  177.61&  -9.77&  19952.&  0.661&  12/04/91&  13:56:20&   0.335\\
   J0500+2529     & 177.18& -10.28&  178.05& -10.50&  11975.&  0.882&  02/21/97&  15:26:49&   0.335\\
   J0510+5545     & 153.99&   9.42&  153.40&   9.68&  12776.&  0.638&  04/09/96&  08:43:27&   0.134\\
   J0510+5545     & 153.99&   9.42&  154.79&   9.37&  23436.&  0.795&  04/29/94&  15:50:10&   0.134\\
   J0520+2556     & 179.65&  -6.40&  179.59&  -7.34&  18127.&  0.942&  10/20/95&  07:32:59&   0.416\\
   J0521+2147     & 183.08&  -8.43&  183.37&  -7.98&  31040.&  0.530&  04/28/91&  23:43:14&   0.378\\
   J0530+1323     & 191.37& -11.01&  191.90& -10.46&  11227.&  0.753&  08/11/95&  16:12:40&   0.352\\
   J0530+1323     & 191.37& -11.01&  192.13& -11.52&  17478.&  0.903&  06/13/91&  08:55:00&   0.352\\
   J0534+2200     & 184.56&  -5.78&  184.32&  -5.90&  10507.&  0.269&  12/13/93&  10:16:58&   0.441\\
   J0534+2200     & 184.56&  -5.78&  184.33&  -5.51&  22733.&  0.353&  06/12/91&  17:32:46&   0.441\\
   J0534+2200     & 184.56&  -5.78&  184.45&  -5.81&  26668.&  0.112&  05/14/99&  00:54:53&   0.441\\
   J0534+2200     & 184.56&  -5.78&  184.45&  -5.99&  14390.&  0.237&  08/25/96&  13:02:24&   0.441\\
   J0534+2200     & 184.56&  -5.78&  184.50&  -5.95&  20975.&  0.175&  12/08/93&  18:40:03&   0.441\\
   J0534+2200     & 184.56&  -5.78&  184.52&  -5.89& 101272.&  0.122&  08/12/92&  22:59:43&   0.441\\
   J0534+2200     & 184.56&  -5.78&  184.55&  -5.96&  12828.&  0.177&  03/21/95&  05:01:49&   0.441\\
   J0534+2200     & 184.56&  -5.78&  184.61&  -5.73&  29156.&  0.071&  04/24/91&  21:57:50&   0.441\\
   J0534+2200     & 184.56&  -5.78&  184.63&  -5.59&  29591.&  0.202&  02/11/94&  21:10:04&   0.441\\
   J0534+2200     & 184.56&  -5.78&  184.67&  -5.85& 117397.&  0.128&  09/15/92&  13:41:36&   0.441\\
   J0542+2610     & 182.02&  -1.99&  182.62&  -1.48&  12418.&  0.789&  02/19/97&  13:25:25&   0.577\\
   J0546+3948     & 170.75&   5.74&  170.65&   5.32&  17104.&  0.432&  08/13/92&  03:41:17&   0.241\\
   J0617+2238     & 189.00&   3.05&  188.73&   2.67&  18595.&  0.465&  08/17/96&  09:24:01&   0.453\\
   J0617+2238     & 189.00&   3.05&  189.35&   3.71&  21052.&  0.745&  02/11/94&  05:14:34&   0.453\\
   J0617+2238     & 189.00&   3.05&  189.37&   3.63&  15826.&  0.687&  08/24/96&  22:41:47&   0.453\\
   J0622$-$1139     & 220.16& -11.69&  219.33& -11.38&  11162.&  0.873&  05/15/95&  10:09:35&   0.132\\
   J0628+1847     & 193.66&   3.64&  194.35&   4.14&  10631.&  0.846&  03/05/97&  00:23:34&   0.350\\
   J0633+1751     & 195.13&   4.27&  194.35&   4.14&  10631.&  0.794&  03/05/97&  00:23:34&   0.296\\
   J0633+1751     & 195.13&   4.27&  194.39&   4.70&  24204.&  0.855&  07/08/98&  13:08:44&   0.296\\
   J0633+1751     & 195.13&   4.27&  194.66&   4.64&  11473.&  0.597&  04/27/00&  07:41:15&   0.296\\
   J0633+1751     & 195.13&   4.27&  194.88&   3.74&  10503.&  0.588&  04/08/95&  03:48:15&   0.296\\
   J0633+1751     & 195.13&   4.27&  195.04&   4.04&  22557.&  0.247&  07/11/98&  21:17:41&   0.296\\
   J0633+1751     & 195.13&   4.27&  195.16&   4.42&  11030.&  0.153&  06/14/91&  14:09:37&   0.296\\
   J0633+1751     & 195.13&   4.27&  195.26&   4.41&  23705.&  0.191&  03/11/95&  08:34:48&   0.296\\
   J0633+1751     & 195.13&   4.27&  195.38&   4.34&  12257.&  0.256&  12/11/93&  17:57:35&   0.296\\
   J0633+1751     & 195.13&   4.27&  195.40&   3.75&  15005.&  0.586&  08/22/94&  03:12:54&   0.296\\
   J0633+1751     & 195.13&   4.27&  195.56&   4.16&  11291.&  0.441&  05/08/00&  05:55:51&   0.296\\
   J0812$-$0646     & 228.64&  14.62&  229.34&  14.57&  10330.&  0.677&  11/04/92&  10:09:52&   0.017\\
   J0827$-$4247     & 260.84&  -2.46&  261.82&  -2.41&  21332.&  0.984&  09/13/94&  21:23:34&   0.381\\
   J0834$-$4511     & 263.55&  -2.79&  262.76&  -3.25&  20763.&  0.910&  09/20/94&  01:32:32&   0.400\\
   J0834$-$4511     & 263.55&  -2.79&  263.03&  -3.22&  16192.&  0.672&  09/04/91&  02:02:34&   0.400\\
   J0834$-$4511     & 263.55&  -2.79&  263.35&  -2.81&  13577.&  0.199&  09/09/94&  08:38:48&   0.400\\
   J0834$-$4511     & 263.55&  -2.79&  263.55&  -2.91&  12719.&  0.119&  09/07/94&  12:58:12&   0.400\\
   J0834$-$4511     & 263.55&  -2.79&  263.97&  -2.80&  11914.&  0.422&  09/10/94&  12:40:20&   0.400\\
   J0848$-$4429     & 264.50&  -0.46&  264.47&   0.50&  16475.&  0.956&  09/02/91&  23:26:25&   0.457\\
   J0848$-$4429     & 264.50&  -0.46&  265.16&  -0.68&  32073.&  0.699&  05/15/91&  22:52:07&   0.457\\
   J0859$-$4257     & 264.57&   2.01&  264.72&   2.38&  15629.&  0.401&  05/11/91&  19:26:28&   0.357\\
   J1013$-$5915     & 283.93&  -2.34&  283.51&  -1.54&  13675.&  0.899&  04/23/95&  13:55:55&   0.530\\
   J1013$-$5915     & 283.93&  -2.34&  284.67&  -2.74&  26761.&  0.841&  09/03/97&  15:35:11&   0.530\\
   J1014$-$5705     & 282.80&  -0.51&  283.48&  -0.93&  18289.&  0.801&  07/02/92&  13:14:34&   0.620\\
   J1027$-$5817     & 284.94&  -0.52&  285.41&  -0.92&  31274.&  0.617&  10/27/94&  06:49:53&   0.570\\
   J1104+3809     & 179.83&  65.03&  180.28&  64.75&  14184.&  0.335&  09/19/92&  23:51:17&   0.044\\
   J1222+2841     & 197.27&  83.52&  202.54&  83.37&  27305.&  0.620&  04/30/93&  15:50:33&   0.040\\
   J1310$-$0517     & 311.69&  57.25&  310.51&  56.59&  10923.&  0.923&  06/22/91&  21:50:09&   0.106\\
   J1324$-$4314     & 309.52&  19.42&  309.06&  19.34&  10019.&  0.445&  04/04/93&  01:16:42&   0.125\\
   J1329+1708     & 346.29&  76.68&  344.50&  77.35&  10761.&  0.784&  12/31/92&  19:29:20&   0.060\\
   J1409$-$0745     & 333.88&  50.28&  333.71&  49.51&  11424.&  0.783&  12/10/94&  19:03:24&   0.070\\
   J1409$-$0745     & 333.88&  50.28&  333.87&  50.65&  11156.&  0.375&  01/24/93&  12:51:18&   0.070\\
   J1409$-$0745     & 333.88&  50.28&  334.28&  50.29&  10065.&  0.258&  12/05/94&  05:44:53&   0.070\\
   J1409$-$0745     & 333.88&  50.28&  334.49&  49.55&  44605.&  0.827&  01/08/93&  21:51:22&   0.070\\
   J1410$-$6147     & 312.18&  -0.35&  312.23&  -0.67&  13366.&  0.321&  04/19/93&  05:17:53&   0.992\\
   J1410$-$6147     & 312.18&  -0.35&  312.24&   0.50&  12185.&  0.848&  10/26/91&  12:33:21&   0.992\\
   J1410$-$6147     & 312.18&  -0.35&  312.35&  -0.37&  17733.&  0.172&  01/12/94&  20:37:53&   0.992\\
   J1410$-$6147     & 312.18&  -0.35&  312.47&  -0.81&  22667.&  0.539&  01/11/94&  07:27:06&   0.992\\
   J1420$-$6038     & 313.63&   0.37&  313.15&   0.26&  18983.&  0.498&  04/18/98&  11:58:39&   0.915\\
   J1420$-$6038     & 313.63&   0.37&  313.45&   0.31&  15880.&  0.188&  05/03/92&  12:33:01&   0.915\\
   J1616$-$2221     & 353.00&  20.03&  352.68&  19.65&  11109.&  0.485&  08/05/94&  08:57:57&   0.195\\
   J1616$-$2221     & 353.00&  20.03&  353.27&  19.91&  12972.&  0.282&  09/18/93&  20:07:41&   0.195\\
   J1625$-$2955     & 348.82&  13.32&  348.13&  13.74&  10015.&  0.790&  06/19/95&  01:54:24&   0.256\\
   J1625$-$2955     & 348.82&  13.32&  348.81&  13.21&  13204.&  0.107&  04/03/94&  10:28:13&   0.256\\
   J1625$-$2955     & 348.82&  13.32&  349.79&  13.10&  13516.&  0.970&  03/26/94&  15:28:51&   0.256\\
   J1626$-$2519     & 352.14&  16.32&  352.39&  16.38&  14241.&  0.247&  12/13/91&  03:04:21&   0.304\\
   J1626$-$2519     & 352.14&  16.32&  352.43&  16.31&  19395.&  0.279&  09/18/93&  02:49:50&   0.304\\
   J1627$-$2419     & 353.36&  16.71&  352.39&  16.38&  14241.&  0.988&  12/13/91&  03:04:21&   0.368\\
   J1627$-$2419     & 353.36&  16.71&  352.43&  16.31&  19395.&  0.976&  09/18/93&  02:49:50&   0.368\\
   J1627$-$2419     & 353.36&  16.71&  353.53&  15.77&  28949.&  0.956&  06/08/95&  04:58:50&   0.368\\
   J1635+3813     &  61.09&  42.34&   61.47&  42.14&  11029.&  0.345&  09/18/91&  19:20:29&   0.026\\
   J1638$-$2749     & 352.25&  12.59&  351.91&  12.17&  22410.&  0.537&  09/14/93&  13:34:44&   0.315\\
   J1639$-$4702     & 337.75&  -0.15&  338.05&  -0.73&  14247.&  0.653&  07/26/91&  18:20:20&   1.681\\
   J1655$-$4554     & 340.48&  -1.61&  340.16&  -1.65&  11164.&  0.327&  05/06/92&  15:48:19&   1.401\\
   J1655$-$4554     & 340.48&  -1.61&  340.55&  -1.75&  10759.&  0.156&  08/28/92&  07:09:46&   1.401\\
   J1655$-$4554     & 340.48&  -1.61&  341.21&  -2.07&  18407.&  0.867&  07/23/91&  11:11:32&   1.401\\
   J1655$-$4554     & 340.48&  -1.61&  341.39&  -1.80&  28011.&  0.929&  12/18/91&  17:51:54&   1.401\\
   J1710$-$4439     & 343.10&  -2.69&  342.43&  -2.14&  11101.&  0.864&  06/29/95&  19:00:14&   1.094\\
   J1710$-$4439     & 343.10&  -2.69&  342.83&  -2.29&  10826.&  0.483&  03/29/94&  20:13:18&   1.094\\
   J1710$-$4439     & 343.10&  -2.69&  342.86&  -2.79&  10702.&  0.262&  07/17/91&  15:09:51&   1.094\\
   J1710$-$4439     & 343.10&  -2.69&  342.88&  -2.85&  11032.&  0.272&  06/20/95&  04:32:08&   1.094\\
   J1710$-$4439     & 343.10&  -2.69&  343.01&  -2.71&  13525.&  0.097&  08/24/97&  13:45:36&   1.094\\
   J1710$-$4439     & 343.10&  -2.69&  343.19&  -2.70&  15681.&  0.091&  07/14/91&  18:56:13&   1.094\\
   J1710$-$4439     & 343.10&  -2.69&  343.30&  -2.82&  15201.&  0.237&  08/29/93&  19:14:07&   1.094\\
   J1710$-$4439     & 343.10&  -2.69&  343.33&  -2.73&  64255.&  0.232&  06/30/95&  08:58:44&   1.094\\
   J1710$-$4439     & 343.10&  -2.69&  343.88&  -2.18&  35943.&  0.931&  07/09/95&  10:36:09&   1.094\\
   J1714$-$3857     & 348.04&  -0.09&  347.41&  -0.11&  26195.&  0.631&  09/19/93&  14:26:08&   2.127\\
   J1714$-$3857     & 348.04&  -0.09&  347.80&   0.79&  23674.&  0.909&  07/05/95&  07:01:04&   2.127\\
   J1714$-$3857     & 348.04&  -0.09&  348.44&  -0.34&  16439.&  0.477&  07/22/91&  10:12:42&   2.127\\
   J1714$-$3857     & 348.04&  -0.09&  348.64&   0.69&  11625.&  0.980&  06/15/95&  15:33:53&   2.127\\
   J1714$-$3857     & 348.04&  -0.09&  348.66&  -0.76&  16427.&  0.911&  06/13/95&  10:57:53&   2.127\\
   J1717$-$2737     & 357.67&   5.95&  357.80&   6.28&  13575.&  0.353&  12/18/91&  12:08:35&   0.862\\
   J1718$-$3313     & 353.20&   2.56&  353.08&   2.76&  12779.&  0.232&  05/06/93&  19:29:08&   1.567\\
   J1718$-$3313     & 353.20&   2.56&  353.79&   1.81&  15613.&  0.959&  08/19/97&  04:41:17&   1.567\\
   J1733$-$1313     &  12.03&  10.81&   11.46&  11.06&  15200.&  0.611&  12/26/91&  14:05:00&   0.407\\
   J1733$-$1313     &  12.03&  10.81&   12.05&  10.12&  11375.&  0.693&  06/19/94&  05:13:39&   0.407\\
   J1733$-$1313     &  12.03&  10.81&   12.16&  10.79&  23716.&  0.130&  04/19/94&  20:49:42&   0.407\\
   J1734$-$3232     & 355.64&   0.15&  354.83&   0.53&  14198.&  0.898&  06/10/95&  08:22:25&   3.007\\
   J1734$-$3232     & 355.64&   0.15&  355.41&   0.55&  20107.&  0.464&  07/15/91&  16:52:10&   3.007\\
   J1734$-$3232     & 355.64&   0.15&  356.14&   0.05&  14205.&  0.512&  12/16/91&  21:04:56&   3.007\\
   J1734$-$3232     & 355.64&   0.15&  356.31&   0.63&  12567.&  0.824&  08/05/94&  21:18:47&   3.007\\
   J1736$-$2908     & 358.79&   1.56&  358.30&   1.86&  17160.&  0.577&  04/02/94&  12:30:05&   2.156\\
   J1736$-$2908     & 358.79&   1.56&  359.44&   1.15&  23192.&  0.769&  07/20/91&  02:15:11&   2.156\\
   J1736$-$2908     & 358.79&   1.56&  359.58&   1.73&  19839.&  0.806&  08/24/93&  18:51:50&   2.156\\
   J1744$-$0310     &  22.19&  13.42&   22.07&  12.45&  13945.&  0.975&  06/12/94&  00:59:33&   0.262\\
   J1744$-$3011     & 358.85&  -0.52&  358.04&  -0.83&  15582.&  0.872&  06/14/95&  21:04:39&   3.171\\
   J1744$-$3011     & 358.85&  -0.52&  358.28&  -1.08&  30387.&  0.797&  06/22/93&  05:38:59&   3.171\\
   J1744$-$3011     & 358.85&  -0.52&  358.65&  -0.84&  11108.&  0.378&  06/21/94&  23:04:04&   3.171\\
   J1744$-$3011     & 358.85&  -0.52&  358.66&  -0.61&  18476.&  0.212&  12/26/95&  23:29:41&   3.171\\
   J1744$-$3011     & 358.85&  -0.52&  359.60&  -1.14&  11850.&  0.972&  08/10/97&  09:22:52&   3.171\\
   J1744$-$3934     & 350.81&  -5.38&  351.14&  -6.07&  28359.&  0.763&  04/01/94&  04:23:02&   0.669\\
   J1746$-$1001     &  16.34&   9.64&   17.05&   9.26&  23043.&  0.797&  04/19/94&  20:51:56&   0.454\\
   J1746$-$2851     &   0.11&  -0.04&    0.15&   0.58&  24511.&  0.622&  07/23/91&  04:53:28&   3.532\\
   J1746$-$2851     &   0.11&  -0.04&    0.15&  -0.18&  14456.&  0.141&  08/07/94&  19:57:23&   3.532\\
   J1746$-$2851     &   0.11&  -0.04&    0.18&  -0.11&  10963.&  0.103&  09/03/93&  02:29:35&   3.532\\
   J1746$-$2851     &   0.11&  -0.04&    0.26&  -0.17&  11462.&  0.200&  08/09/93&  04:30:24&   3.532\\
   J1746$-$2851     &   0.11&  -0.04&    0.41&   0.10&  21570.&  0.335&  09/14/93&  22:52:06&   3.532\\
   J1746$-$2851     &   0.11&  -0.04&    0.86&   0.39&  11358.&  0.864&  06/11/95&  04:23:19&   3.532\\
   J1746$-$2851     &   0.11&  -0.04&  359.84&  $-$0.30&  17665.&  0.376&  06/29/93&  08:22:43&   3.532\\
   J1746$-$2851     &   0.11&  -0.04&  359.95&  $-$0.38&  18736.&  0.379&  07/23/91&  17:18:54&   3.532\\
   J1757$-$0711     &  20.30&   8.47&   20.61&   8.68&  11566.&  0.371&  06/12/94&  22:35:13&   0.548\\
   J1800$-$2338     &   6.25&  -0.18&    6.63&  -0.23&  11482.&  0.382&  06/18/95&  13:23:01&   2.967\\
   J1800$-$3955     & 352.45&  -8.43&  352.38&  -8.60&  29552.&  0.188&  06/27/94&  06:49:05&   0.379\\
   J1809$-$2328     &   7.47&  -1.99&    7.09&  -2.15&  15065.&  0.412&  09/22/95&  13:19:51&   2.238\\
   J1809$-$2328     &   7.47&  -1.99&    7.33&  -2.14&  15083.&  0.207&  07/16/91&  23:44:22&   2.238\\
   J1809$-$2328     &   7.47&  -1.99&    7.84&  -2.83&  10407.&  0.920&  08/26/93&  19:48:46&   2.238\\
   J1809$-$2328     &   7.47&  -1.99&    8.02&  -2.53&  12446.&  0.767&  03/30/93&  11:42:13&   2.238\\
   J1812$-$1316     &  16.70&   2.39&   16.00&   2.51&  23715.&  0.709&  03/29/93&  18:36:58&   1.725\\
   J1823$-$1314     &  17.94&   0.14&   17.44&  -0.50&  11321.&  0.810&  07/02/94&  09:33:12&   2.681\\
   J1823$-$1314     &  17.94&   0.14&   17.89&  -0.27&  21167.&  0.415&  06/14/94&  08:53:49&   2.681\\
   J1823$-$1314     &  17.94&   0.14&   18.10&  -0.37&  14931.&  0.534&  06/13/94&  15:32:37&   2.681\\
   J1823$-$1314     &  17.94&   0.14&   18.12&   0.48&  13397.&  0.385&  09/17/93&  07:36:49&   2.681\\
   J1824$-$1514     &  16.37&  -1.16&   15.87&  -1.31&  21376.&  0.526&  12/15/91&  03:24:59&   2.489\\
   J1824$-$1514     &  16.37&  -1.16&   16.27&  -0.83&  20998.&  0.344&  07/18/94&  21:47:57&   2.489\\
   J1824$-$1514     &  16.37&  -1.16&   16.79&  -1.25&  16468.&  0.431&  09/18/93&  16:56:10&   2.489\\
   J1824$-$1514     &  16.37&  -1.16&   16.99&  -1.87&  17197.&  0.944&  04/20/94&  21:17:46&   2.489\\
   J1826$-$1302     &  18.47&  -0.44&   17.89&  -0.27&  21167.&  0.607&  06/14/94&  08:53:49&   2.793\\
   J1826$-$1302     &  18.47&  -0.44&   18.10&  -0.37&  14931.&  0.377&  06/13/94&  15:32:37&   2.793\\
   J1826$-$1302     &  18.47&  -0.44&   18.12&   0.48&  13397.&  0.986&  09/17/93&  07:36:49&   2.793\\
   J1826$-$1302     &  18.47&  -0.44&   18.71&  -0.61&  10302.&  0.291&  02/15/92&  18:43:03&   2.793\\
   J1826$-$1302     &  18.47&  -0.44&   19.45&  -0.50&  10810.&  0.980&  11/07/91&  00:56:16&   2.793\\
   J1837$-$0423     &  27.44&   1.06&   27.02&   0.96&  32150.&  0.434&  06/19/94&  19:37:18&   1.560\\
   J1837$-$0423     &  27.44&   1.06&   27.79&   0.32&  19447.&  0.819&  08/18/91&  09:35:09&   1.560\\
   J1837$-$0606     &  25.86&   0.40&   25.05&   0.47&  41742.&  0.815&  04/25/94&  13:31:39&   2.082\\
   J1837$-$0606     &  25.86&   0.40&   25.08&   0.21&  12889.&  0.804&  08/05/93&  04:18:36&   2.082\\
   J1837$-$0606     &  25.86&   0.40&   25.29&   0.25&  10303.&  0.590&  06/30/95&  01:10:13&   2.082\\
   J1837$-$0606     &  25.86&   0.40&   25.35&  -0.28&  14864.&  0.843&  02/15/92&  21:33:20&   2.082\\
   J1837$-$0606     &  25.86&   0.40&   25.51&   0.44&  20056.&  0.347&  04/21/94&  16:14:34&   2.082\\
   J1837$-$0606     &  25.86&   0.40&   25.90&  -0.33&  35420.&  0.727&  11/01/91&  16:26:39&   2.082\\
   J1837$-$0606     &  25.86&   0.40&   26.59&   0.19&  10863.&  0.764&  12/25/95&  22:53:44&   2.082\\
   J1847$-$3219     &   3.21& -13.37&    3.72& -13.31&  23065.&  0.498&  06/13/95&  08:10:28&   0.246\\
   J1856+0114     &  34.60&  -0.54&   33.82&  -0.49&  19294.&  0.782&  02/16/92&  14:43:38&   1.273\\
   J1856+0114     &  34.60&  -0.54&   33.82&  -0.69&  20271.&  0.798&  02/16/92&  01:12:29&   1.273\\
   J1856+0114     &  34.60&  -0.54&   33.86&  -0.55&  12031.&  0.741&  10/07/95&  15:56:28&   1.273\\
   J1856+0114     &  34.60&  -0.54&   34.61&  -0.19&  11616.&  0.346&  10/08/95&  21:18:32&   1.273\\
   J1856+0114     &  34.60&  -0.54&   34.92&   0.00&  10226.&  0.633&  10/08/95&  08:25:09&   1.273\\
   J1858$-$2137     &  14.21& -11.15&   14.33& -10.27&  15269.&  0.890&  09/02/93&  14:52:02&   0.284\\
   J1903+0550     &  39.52&  -0.05&   38.94&  -0.44&  11318.&  0.702&  11/06/91&  19:04:16&   1.183\\
   J1903+0550     &  39.52&  -0.05&   40.03&  -0.30&  11140.&  0.566&  02/07/92&  14:16:36&   1.183\\
   J1904$-$1124     &  24.22&  -8.12&   23.57&  -8.03&  11701.&  0.652&  02/19/92&  12:04:27&   0.327\\
   J1935$-$4022     & 358.65& -25.23&  358.20& -24.43&  15977.&  0.895&  04/03/94&  18:49:48&   0.091\\
   J1958+2909     &  66.23&  -0.16&   65.81&  -0.99&  33888.&  0.935&  10/29/96&  00:19:12&   0.683\\
   J2016+3657     &  74.76&   0.98&   75.49&   1.14&  10366.&  0.744&  03/15/93&  01:02:16&   0.767\\
   J2020+4017     &  78.05&   2.08&   77.79&   2.27&  14319.&  0.318&  05/31/91&  13:20:18&   0.820\\
   J2020+4017     &  78.05&   2.08&   78.81&   1.58&  16945.&  0.910&  12/13/92&  22:51:58&   0.820\\
   J2021+3716     &  75.58&   0.33&   75.49&   1.14&  10366.&  0.811&  03/15/93&  01:02:16&   0.824\\
   J2022+4317     &  80.63&   3.62&   80.05&   3.22&  11593.&  0.701&  09/23/93&  09:54:56&   0.672\\
   J2033+4118     &  80.27&   0.73&   80.52&   1.46&  11256.&  0.768&  05/30/94&  14:33:03&   0.980\\
   J2033+4118     &  80.27&   0.73&   80.60&   1.00&  14884.&  0.429&  06/06/91&  01:32:11&   0.980\\
   J2100+6012     &  97.76&   9.16&   98.08&   9.27&  12137.&  0.333&  12/20/92&  15:17:15&   0.176\\
   J2158$-$3023     &  17.73& -52.25&   17.42& -52.35&  11138.&  0.210&  02/20/93&  01:46:50&   0.024\\
   J2158$-$3023     &  17.73& -52.25&   17.65& -52.14&  11183.&  0.115&  05/22/96&  11:06:37&   0.024\\
   J2158$-$3023     &  17.73& -52.25&   17.92& -52.18&  11069.&  0.137&  01/07/98&  20:29:19&   0.024\\
   J2219$-$7941     & 310.64& -35.06&  310.12& -34.42&  15047.&  0.771&  06/11/96&  03:35:47&   0.036\\
 \enddata

\end{deluxetable}

\clearpage

\begin{deluxetable}{rrrrrrr}
\tabletypesize{\footnotesize}
\tablecaption{$E>10$ GeV Events from the Crab. \label{tbl-5}}
\tablewidth{0pt}
\tablehead{
\colhead{Date} & \colhead{Time}   & \colhead{Energy}  & \colhead{Energy} & \colhead{Galactic}   &
\colhead{Galactic} & \colhead{Pulsar}\\
& &  \colhead{(MeV)} & \colhead{Uncertainty} &  \colhead{Latitude} & \colhead{Longitude} & \colhead{Phase\tablenotemark{a}}}
\startdata
4/24/1991 &	21:57:50.485 &	29156 &	5870 &	-5.73&	184.61& 0.697\\
6/12/1991 &	17:32:46.514 &	22733	&     5328 &	-5.51&	184.33& 0.353\\
8/12/1992&	22:59:43.472&	101272&	52495&	-5.89&	184.52& 0.100\\
9/15/1992&	13:41:36.138&	117397&	56213	&-5.85	&184.67& 0.070\\
12/8/1993&	18:40:03.898&	20975	&5405	&-5.95	&184.50& 0.079\\
12/13/1993&	10:16:58.551&	10507	&1925	&-5.90&	184.32& 0.729\\
2/11/1994&	21:10:04.761&	29591	&6310	&-5.59&	184.63& 0.522\\
3/21/1995&	05:01:49.000&	12828	&2205	&-5.96&	184.55& 0.345\\
8/25/1996&	13:02:24.686&	14390	&2487	&-5.99&	184.45& 0.754\\
5/14/1999&	00:54:53.468&	26668	&7943	&-5.81&	184.45& 0.470
 \enddata

\tablenotetext{a}{The main radio pulse appears at phase 0.10}

\end{deluxetable}

\clearpage

\begin{deluxetable}{rrrrrrr}
\tabletypesize{\footnotesize}
\tablecaption{$E>10$ GeV Events from Geminga. \label{tbl-6}}
\tablewidth{0pt}
\tablehead{
\colhead{Date} & \colhead{Time}   & \colhead{Energy}  & \colhead{Energy} & \colhead{Galactic}   &
\colhead{Galactic} & \colhead{Pulsar}\\
& &  \colhead{(MeV)} & \colhead{Uncertainty} &  \colhead{Latitude} & \colhead{Longitude} & \colhead{Phase\tablenotemark{a}}}
\startdata
6/14/1991&	14:09:37.395&	11030&	1940&	4.42&	195.16& 0.599\\
12/11/1993&	17:57:35.652&	12257&	1810&	4.35&	195.38& 0.583\\
8/22/1994&	03:12:54.489&	15005&	3614&	3.75&	195.40& 0.003\\
3/11/1995&	08:34:48.482&	23705&	5284&	4.41&	195.26& 0.578\\
4/8/1995&	03:48:15.999&	10503&	2373&	3.74&	194.88& 0.470\\
3/5/1997&	00:23:34.124&	10631&	1974&	4.14&	194.35& 0.620\\
7/8/1998&	13:08:44.888&	24204&	4868&	4.70&	194.39& 0.580\\
7/11/1998&	21:17:41.710&	22557&	4407&	4.04&	195.04& 0.700\\
4/27/2000&	07:41:15.494&	11473&	2014&	4.64&	194.66& 0.370\\
5/8/2000&	05:55:51.720&	11291&	1984&	4.16&	195.56& 0.280
 \enddata

\tablenotetext{a}{There is no clearly-defined radio phase.}

\end{deluxetable}

\clearpage

\begin{deluxetable}{rrrrrrr}
\tabletypesize{\footnotesize}
\tablecaption{$E>10$ GeV Events from Vela. \label{tbl-7}}
\tablewidth{0pt}
\tablehead{
\colhead{Date} & \colhead{Time}   & \colhead{Energy}  & \colhead{Energy} & \colhead{Galactic}   &
\colhead{Galactic} & \colhead{Pulsar}\\
& &  \colhead{(MeV)} & \colhead{Uncertainty} &  \colhead{Latitude} & \colhead{Longitude} & \colhead{Phase\tablenotemark{a}}}
\startdata
9/4/1991&	02:02:34.878&	16192&	2700&	-3.22&	263.03& 0.464\\
9/7/1994&	12:58:12.854&	12719&	5220&	-2.91&	263.56& 0.305\\
9/9/1994&	08:38:48.784&	13577&	2065&	-2.81&	263.35& 0.465\\
9/10/1994&	12:40:20.170&	11914&	1754&	-2.80&	263.97& 0.053\\
 \enddata

\tablenotetext{a}{The radio pulse appears at phase 0.95.}

\end{deluxetable}

\clearpage

\begin{deluxetable}{rrrrrrr}
\tabletypesize{\footnotesize}
\tablecaption{$E>10$ GeV Events from PSR B1706$-$44. \label{tbl-8}}
\tablewidth{0pt}
\tablehead{
\colhead{Date} & \colhead{Time}   & \colhead{Energy}  & \colhead{Energy} & \colhead{Galactic}   &
\colhead{Galactic} & \colhead{Pulsar}\\
& &  \colhead{(MeV)} & \colhead{Uncertainty} &  \colhead{Latitude} & \colhead{Longitude} & \colhead{Phase\tablenotemark{a}}}
\startdata
7/14/1991&	18:56:13.768&	15681&	3630&	-2.70&	343.19& 0.248\\
7/17/1991&	15:09:51.132&	10702&	2062&	-2.79&	342.86& 0.259\\
8/29/1993&	19:14:07.261&	15201&	2630&	-2.82&	343.30& 0.480\\
3/29/1994&	20:13:18.277&	10826&	2067&	-2.29&	342.83& 0.257\\
6/20/1995&	04:32:08.336&	11032&	1983&	-2.85&	342.88& 0.295\\
6/29/1995&	19:00:14.406&	11101&	2310&	-2.14&	342.43& 0.809\\
6/30/1995&	08:58:44.104&	64255&	33741&	-2.73&	343.33& 0.525\\
7/9/1995&	10:36:09.665&	35943&	14368&	-2.18&	343.88& 0.902\\
8/24/1997&	13:45:36.126&	13525&	2907&	-2.71&	343.01& 0.223
 \enddata

\tablenotetext{a}{The radio pulse appears at phase 0.0.}

\end{deluxetable}

\clearpage

\begin{deluxetable}{rrrrrrr}
\tabletypesize{\footnotesize}
\tablecaption{$E>10$ GeV Events from PSR B1951+32. \label{tbl-9}}
\tablewidth{0pt}
\tablehead{
\colhead{Date} & \colhead{Time}   & \colhead{Energy}  & \colhead{Energy} & \colhead{Galactic}   &
\colhead{Galactic} & \colhead{Pulsar}\\
& &  \colhead{(MeV)} & \colhead{Uncertainty} &  \colhead{Latitude} & \colhead{Longitude} & \colhead{Phase\tablenotemark{a}}}
\startdata
3/21/1993&	19:05:57.003&	17566&	4391&	2.61&	68.86& 0.613\\
7/9/1994&	22:18:26.278&	13829&	2320&	3.45&	68.52& 0.622
 \enddata

\tablenotetext{a}{The radio pulse appears at phase 0.0.}

\end{deluxetable}


\end{document}